\documentclass{PoS}

\usepackage{slashed}

\newcommand{\pslash}{\ensuremath{\slashed{p}}}

\newcommand{\Drsl}{\ensuremath{\stackrel{\rightarrow}{\slashed{D}}}}
\newcommand{\Aslash}{\ensuremath{\slashed{A}}}

\newcommand{\half}{{\textstyle \frac{1}{2}}}

\newcommand{\ggCF}{\frac{g^2\, C_F}{16 \pi^2} }
\newcommand{\ggMFCF}{\frac{g^2_{MF}\, C_F}{16 \pi^2} }

\newcommand{\be}{\begin{equation}}
\newcommand{\ee}{\end{equation}}

\newcommand{\bea}{\begin{eqnarray}}
\newcommand{\eea}{\end{eqnarray}}

\leftline{{\normalsize DESY 08-126} }
\leftline{{\normalsize Edinburgh 2008/29}} 
\leftline{{\normalsize Leipzig LU-ITP 2008/03}} 
\leftline{{\normalsize Liverpool LTH 803}}
\vspace{1cm}

\title{Clover improvement for stout-smeared 2+1 flavour 
SLiNC fermions: perturbative results }

\ShortTitle{SLiNC fermions: perturbation theory}

\author{R.~Horsley$^{a}$, \speaker{H.~Perlt}$^{,b}$, P.E.L.~Rakow$^{c}$, G.~Schierholz$^{d}$, and  A.~Schiller$^{b}$ \\
~\\

$^{a}$ School of Physics and Astronomy, University of Edinburgh, Edinburgh EH9 3JZ, UK \\
$^{b}$ Institut f\"ur Theoretische Physik, Universit\"at Leipzig, D-04109 Leipzig, Germany \\
$^{c}$ Theoretical Physics Division, Department of Mathematical Sciences, University of Liverpool,  Liverpool L69 3BX, UK \\
$^{d}$Deutsches Elektronen-Synchrotron DESY, D-22603 Hamburg, Germany\\
}

\abstract{For the {\bf S}tout {\bf Li}nk {\bf N}on-perturbative {\bf C}lover (SLiNC) action we determine in one-loop lattice
perturbation theory the critical hopping parameter $\kappa_c$ and the clover parameter $c_{SW}$ which is needed
for $\mathcal{O}(a)$ improvement. Performing this calculation off-shell we are also able to compute  the
non gauge invariant quark field improvement coefficient $c_{NGI}$ . Additionally, we present first results for
the renormalization factors of the scalar, pseudoscalar, vector and axial vector currents. We discuss
 mean field improvement for the SLiNC action.}

\FullConference{The XXVI International Symposium on Lattice Field Theory\\
		 July 14-19 2008\\
		 Williamsburg, Virginia, USA}

\begin{document}


\section{Introduction}

Simulations of Wilson-type fermions at realistic quark masses require an
improved action with good chiral properties and scaling behavior. 
A systematic improvement scheme that removes discretization errors order by
order in the lattice spacing $a$ has been proposed by
Symanzik~\cite{Symanzik:1983dc} and developed for on-shell quantities
in~\cite{Luscher:1984xn,Sheikholeslami:1985ij}. $\mathcal{O}(a)$ 
improvement of the Wilson fermion action is achieved by complementing
it with the so-called clover term~\cite{Sheikholeslami:1985ij}, provided the
associated clover coefficient is tuned properly.   

The focus of this contribution is to determine the clover coefficient and the
additive mass renormalization for plaquette and Symanzik improved gauge action
and stout link clover fermions in one-loop lattice perturbation theory.
We correct earlier results published in~\cite{Horsley:2007fw} and introduce a modified mean
field improvement for partially smeared links.
A detailed discussion can be found in~\cite{Horsley:2008ap}. Additionally, 
in this paper we
present first results for the one-loop renormalization factors of the
scalar, pseudoscalar, vector and axial vector currents with the chosen
action. First non-perturbative results obtained with this action
are presented in~\cite{Horsley:2008}.

The Symanzik improved gauge action reads~\cite{Symanzik:1983dc}
\begin{equation}
  S_G^{\rm Sym}= \frac{6}{g^2} \,\,\left\{c_0
  \sum_{\rm Plaquette} \frac{1}{3}\, {\rm Re\, Tr\,}(1-U_{\rm Plaquette})
+ \, c_1 \sum_{\rm Rectangle} \frac{1}{3}\,{\rm Re \, Tr\,}(1- U_{\rm
  Rectangle})\right\}
  \label{SG}
\end{equation}
with $c_0+8c_1=1$ and 
\begin{equation}
  c_0=\frac{5}{3}\,, \quad c_1=-\frac{1}{12}\,.
\end{equation}

Clover fermions have the action for each quark flavor~\cite{Sheikholeslami:1985ij}
\begin{eqnarray} 
  S_F &=& a^4\, \sum_x \Big\{ - \frac{1}{2a} \, \left[\bar{\psi}(x)
  \widetilde U_\mu(x)\,(1-\gamma_\mu)\, \psi(x+a\hat{\mu})
  + \, \bar{\psi}(x) \widetilde U_\mu^\dagger(x-a\hat{\mu})\,(1+\gamma_\mu)\,
  \psi(x-a\hat{\mu})\right] 
\nonumber
\\
&&
  + \, \frac{1}{a}\, (4 + a m_0 +a m)\, \bar{\psi}(x)\psi(x)
  - c_{SW}\, g\, \frac{a}{4}\, \bar{\psi}(x)\,
  \sigma_{\mu\nu} F_{\mu\nu}(x)\, \psi(x) \Big\} \,,
\label{SF}
\end{eqnarray}
where
\begin{equation}
  am_0=\frac{1}{2\kappa_c} - 4 \,,
  \label{kc}
\end{equation}
$\kappa_c$ being the critical hopping parameter, is the additive mass
renormalization term, and $F_{\mu\nu}(x)$ is the field strength tensor in
clover form with
$\sigma_{\mu\nu}=(i/2)\,(\gamma_\mu\gamma_\nu-\gamma_\nu\gamma_\mu)$.  
We consider a version of clover fermions in which we do not smear links
in the clover term, but
the link variables $U_\mu$ in the next neighbor terms have been replaced by
(uniterated) stout links~\cite{Morningstar:2003gk}  
\begin{equation}
  \widetilde{U}_\mu(x) = e^{i\, Q_\mu(x)} \, U_\mu(x)
  \label{Ustout}
\end{equation}
with
\begin{equation}
  Q_\mu(x) =\frac{\omega}{2\,i} \Big[V_\mu(x) U_\mu^\dagger(x) -
  U_\mu(x)V_\mu^\dagger(x) 
-\frac{1}{3} {\rm Tr} \,\left(V_\mu(x)
  U_\mu^\dagger(x) -  U_\mu(x)V_\mu^\dagger(x)\right)\Big] \, .
\end{equation}
$V_\mu(x)$ denotes the sum over all staples associated with the link and
$\omega$ is a tunable weight factor. 
Stout smearing is preferred because (\ref{Ustout}) is expandable as 
a power series in $g^2$, so we can use perturbation theory. Many other
forms of smearing do not have this nice property. 
Because both the unit matrix and the $\gamma_\mu$ terms are smeared, 
each link is still a projection operator in the Dirac spin index. 

The reason for not smearing the clover term is that we want to keep 
the physical extent in lattice units of the fermion matrix small which is relevant 
for non-perturbative calculations.
In that respect we refer to these fermions as SLiNC fermions, from the phrase
{\bf S}tout {\bf Li}nk {\bf N}on-perturbative {\bf C}lover. 
The improvement coefficient $c_{SW}$ as well as the additive mass
renormalization $am_0$ are associated with the chiral limit. So we will carry
out the calculations for massless quarks, which simplifies things, though
it means that we cannot present values for the mass dependent corrections.

In perturbation theory 
\begin{equation}
  c_{SW}=1 + g^2 \, c_{SW}^{(1)} + {\mathcal{O}(g^4)}\,.
  \label{csw}
\end{equation}
The one-loop coefficient $c_{SW}^{(1)}$ has been computed for the plaquette
action using twisted antiperiodic boundary conditions~\cite{Wohlert:1987rf}
and Schr\"odinger functional methods~\cite{Luscher:1996vw}. Moreover, using
conventional perturbation theory, Aoki and Kuramashi~\cite{Aoki:2003sj} have
computed $c_{SW}^{(1)}$ for certain improved gauge actions. All calculations
were performed for non-smeared links and limited to on-shell quantities.

We extend previous calculations of $c_{SW}^{(1)}$ to include stout links. This
is done by computing the one-loop correction to the off-shell quark-quark-gluon
three-point function. The improvement of the action is not sufficient to remove
discretization errors from Green functions. To achieve this, one must
also improve the quark fields~\cite{Horsley:2008ap}
\begin{equation}
  \psi_{\star}(x)=\left(1 + a \,c_D \Drsl + a \,i\,g\,\,c_{NGI} \Aslash(x) \right) \,\psi(x)\,,
  \label{imppsi}
\end{equation}
where the improvement factor $c_{NGI}$ has been introduced by~\cite{Martinelli:2001ak}
and has the perturbative expansion
\begin{equation}
  c_{NGI}=g^2\,c_{NGI}^{(1)} + {\mathcal{O}(g^4)}\,.
  \label{cNGI}
\end{equation}
A detailed discussion of the implications of
off-shell improvement is given in~\cite{Horsley:2008ap}. In this contribution
we concentrate on the on-shell relevant parameters $c_{SW}$ and $\kappa_c$.

\section{Off-shell improvement}

It is known~\cite{Aoki:2003sj} that the one-loop contribution of the Sheikoleslami-Wohlert 
coefficient in conventional perturbation theory can be determined using the quark-quark-gluon 
vertex $\Lambda_\mu(p_1,p_2,c_{SW})$ sandwiched between {\sl on-shell} quark states.
$p_1$ ($p_2$) denotes the incoming (outgoing) quark momentum.
In general that vertex is an {\sl amputated} three-point Green function.

Let us  look at the ${\mathcal{O}}(a)$ expansion of tree-level 
$\Lambda^{(0)}_\mu(p_1,p_2,c_{SW})$ which is derived from action (\ref{SF})
\begin{eqnarray}
  \Lambda^{(0)}_\mu(p_1,p_2,c_{SW}) &=& -i\, g  \,\gamma_\mu -g\, \half \, a\, 
  {\bf 1} (p_1 + p_2)_\mu 
+ \,c_{SW} \,i\, g\, \half \, a \,\sigma_{\mu \alpha} (p_1 -p_2)_\alpha
 +\,\mathcal{O}(a^2)\,.
  \label{treevertex}
\end{eqnarray}
It is obvious from (\ref{treevertex}) that a one-loop calculation of the quark-quark-gluon
vertex provides the needed relation to compute $c_{SW}$ in one -loop also. 

The {\sl off-shell} improvement condition states that the
{\sl non-amputated} improved quark-quark-gluon Green function $G_{\star \mu}(p_1,p_2,q)$ 
has to be free of $\mathcal{O}(a)$ terms in one-loop accuracy. The relation
between the amputated and non-amputated Green functions is
\begin{eqnarray}
G_\mu(p_1,p_2,q) &=& S(p_2)\, \Lambda_\nu(p_1,p_2,q,c_{SW}^{(1)})\, S(p_1)\, K_{\nu\mu}(q)\,.
  \label{nonamp}
\end{eqnarray}
$K_{\nu\mu}(q)$ denotes the full gluon propagator which is $\mathcal{O}(a)$-improved 
already, $S(p)$ the corresponding quark propagator.
Using the improved quark fields one obtains  the following off-shell improvement
condition in momentum space (for details of the derivation see~\cite{Horsley:2008ap})
\begin{eqnarray}
  \Lambda_{\mu}(p_1,p_2,q,c_{SW}^{(1)})&=&\Lambda_{\star \mu}(p_1,p_2,q)+ 
        a \, g^3  c_{NGI}^{(1)} (\pslash_2 \, \gamma_\mu +\gamma_\mu\, \pslash_1)
  \nonumber\\
  & &  -\, \frac{a}{2}\,i\, \pslash_2 \, \frac{\Sigma_2(p_2)}{\Sigma_1(p_2)}\, 
     \Lambda_{\star\mu}(p_1,p_2,q)
     -\frac{a}{2}\,\Lambda_{\star\mu}(p_1,p_2,q)\,  i\, \pslash_1 \, \frac{\Sigma_2(p_1)}{\Sigma_1(p_1)}
  \,,
\label{impcond}
\end{eqnarray}
where the improved three-point function $\Lambda_{\star \mu}(p_1,p_2,q)$ is free of $\mathcal{O}(a)$ terms. 
In (\ref{impcond}) the
quantities $\Sigma_i(p)$ are the corresponding contributions to the quark self energy
\begin{equation}
  \Sigma(p)= \frac{1}{a} \Sigma_0 + i \, \pslash \, \Sigma_1(p) + \frac{a \, p^2}{2} \Sigma_2(p)\,.
\label{Sigma}
\end{equation}

\section{Results for improvement parameters}

The calculation has been performed in general covariant gauge. We use a combination of
symbolic and numeric routines. 

The anticipated general structure for the amputated three-point function at one-loop is
\begin{eqnarray}
  \Lambda_\mu(p_1,p_2,q)&=& \Lambda^{{\overline{MS}}}_\mu(p_1,p_2,q)
  +A_{\rm lat}\,i\,\frac{g^3}{16\pi^2}\,\gamma_\mu
  \nonumber\\
  & & \hspace{-5mm} + \, B_{\rm lat}\,\frac{a}{2}\,\frac{g^3}{16\pi^2}\,\left(\pslash_2\,\gamma_\mu
  +\gamma_\mu\,\pslash_1\right)
+ C_{\rm lat}\,i\,\frac{a}{2}\,\frac{g^3}{16\pi^2}\,\sigma_{\mu\alpha}\,q_\alpha \,.
\label{Lam}
\end{eqnarray}
$\Lambda^{{\overline{MS}}}_\mu(p_1,p_2,q)$ is the universal part of the three-point function,
independent of the chosen gauge action, computed in the $\overline{MS}$-scheme. It is
given in a complete symbolic form in~\cite{Horsley:2008ap}. 

If we insert (\ref{Lam}) into the off-shell
improvement relation (\ref{impcond}) we get the following conditions that all terms of order 
$\mathcal{O}(ag^3)$ have to vanish
\begin{eqnarray}
\left(c_{SW}^{(1)} - \frac{C_{\rm lat}}{16\pi^2}\right)\,\sigma_{\mu\alpha}\,q_\alpha &=& 0\,,
  \label{cSWcond}\\
\left(c_{NGI}^{(1)} - \frac{1}{32\pi^2}\,\left(A_{\rm lat}-B_{\rm lat}-\Sigma_{21}\right)\right)
     \left(\pslash_2 \, \gamma_\mu+ \gamma_\mu\, \pslash_1 \right)
 &=&0\,,
  \label{cNGIcond}
\end{eqnarray}
with $\Sigma_{21}$ defined from (\ref{Sigma}) as
\begin{eqnarray}
  \frac{\Sigma_2(p)}{\Sigma_1(p)}&=&1+\frac{g^2\,C_F}{16\pi^2}\Big((1-\xi)(1-\log(a^2p^2))
+ \Sigma_{21,0} \Big)
  \nonumber\\
  &\equiv&1+\frac{g^2\,C_F}{16\pi^2}\Big((1-\xi)(1-\log(p^2/\mu^2))\Big)
+\frac{g^2}{16\pi^2}\Sigma_{21}
  \label{SigmaWP}
\end{eqnarray}
and
\begin{equation}
  \Sigma_{21}=C_F\,\left( -(1-\xi)\log(a^2\mu^2)+\Sigma_{21,0} \right)\,.
  \label{SigmaWPr}
\end{equation}
The constant $\Sigma_{21,0}$ depends on the chosen lattice action. Inserting the  numbers 
for the Symanzik action we get the following results for the clover improvement coefficient
\begin{eqnarray}
c_{SW}^{(1)}&=&C_F\,\Big(0.116185 + 0.828129\,\omega 
 - 2.455080\,\omega^2\Big)
  \nonumber \\
  &+& N_c\,\Big(0.013777 + 0.015905\,\omega 
- 0.321899\,\omega^2\Big)\,,
  \label{cswSym}
\end{eqnarray}
and for the off-shell quark field improvement coefficient
\begin{eqnarray}
  c_{NGI}^{(1)}&=& N_c\,\left(0.002395 - 0.010841\,\omega \right)\,.
  \label{cNGISym}
\end{eqnarray}
For $\omega=0$ the Symanzik result
(\ref{cswSym}) agree, within the accuracy of our calculations,  
with the number quoted in~\cite{Aoki:2003sj}.

The additive mass renormalization is given by
\begin{equation}
  am_0=\ggCF \,\frac{\Sigma_0}{4} \,.
\end{equation}
This leads to the critical hopping parameter $\kappa_c$, at which chiral
symmetry is approximately restored, 
\begin{equation}
  \kappa_c=\frac{1}{8}\left( 1-
  \ggCF \,\frac{\Sigma_0}{4}\right)\,.
  \label{kappac}
\end{equation}
We obtain the following perturbative expression for $\kappa_c$
\begin{eqnarray}
  \kappa_c & =& \frac{1}{8}  \left[ 1   + g^2 \, C_F
  \left( 0.037730 - 0.662090\,\omega +2.668543\,\omega^2
  \right) \right]
  \,.
  \label{kappacSym}
\end{eqnarray}
 $am_0$ can be tuned to zero for admissible values of $\omega$.
Using the smaller possible value we find $\omega=0.088689$ for the Symanzik gauge action
which is not far away from the value $\omega=0.1$ used in our non-perturbative calculations~\cite{Horsley:2008}.

\section{Mean field improvement}

In the mean field approximation we typically assume that the gauge fields on each link are 
independently fluctuating variables, and that we can simply represent the links by an average
value $u_0$. Typical choices for $u_0$ would be to choose $u_0^4$ to be the average plaquette value, 
or to choose $u_0$ to be the average link value in the Landau gauge. 

A natural question is how we should extend the mean field approximation if we employ smearing. 
One possibility is to express everything in terms of two quantities, $u_0$, a mean value 
for the unsmeared link, and $u_S$, a mean value for smeared links~\footnote{We would like to thank 
Colin Morningstar for conversations on this point.}. 
Applying the mean field approximation to SLiNC fermions we find
\begin{equation}
\kappa_c \approx \frac{1}{8 u_S}, \quad c_{SW} \approx \frac{u_S}{u_0^4}\,.
\end{equation}
As a result, we find  mean field improved expressions for $\kappa_c$ and $c_{SW}$ by performing 
the following replacements
\begin{equation}
  \kappa_c(g^2) \rightarrow \kappa_c^{MF}(g_{MF}^2,u_S)=
        \frac{u_S^{\rm pert}(g_{MF}^2)}{u_S}\,\kappa_c(g_{MF}^2)
\label{kappaMF}
\end{equation}
and
\begin{equation}
c_{SW}(g^2) \rightarrow c_{SW}^{MF}(g_{MF}^2,u_S,u_0)
 =\frac{u_S}{u_0^4}\,\frac{u_0^{\rm pert}(g_{MF}^2)^{\,4}}{u_S^{\rm
        pert}(g_{MF}^2)}\,c_{SW}(g_{MF}^2)\,. 
\label{cswMF}
\end{equation}
Here $u_S$ and $u_0$ are the measured smeared and unsmeared links at the given coupling and
$u_S^{\rm pert}$ and $u_0^{\rm pert}$ denote the corresponding expressions in lattice perturbation
theory.

We will use $u_S^{\rm pert}$ derived from the smeared perturbative plaquette $P_S$ 
\begin{equation}
  u_S^{\rm pert} \equiv P_S^{1/4}. 
\end{equation} 
To one-loop order we have
\begin{equation}
  u_S^{\rm pert}(\omega) = 1 - \ggMFCF\, k_S(\omega)\,,
\end{equation}
where the one-loop contribution  $k_S$ is~\cite{Horsley:2008ap} 
\begin{eqnarray}
  k_S(\omega) &=& \pi^2 \left(  0.732525  -11.394696\,\omega  +   50.245225\,\omega^2 \right)\,.
\label{kS}
\end{eqnarray} 
The unsmeared perturbative value for $u_0^{\rm pert}$ is  $u_0^{\rm pert}=u_S^{\rm pert}(0)$.
Inserting the result (\ref{kS}) into the mean field expressions (\ref{kappaMF}) and
(\ref{cswMF}) we obtain
\begin{eqnarray} 
  \kappa_c^{ MF}   &=&  \frac{1}{8 u_S} 
  \left[1
  +  g_{MF}^2 \,C_F  \,   
  \left( -0.008053 +  0.0500781\,\omega -0.471784\,\omega^2 
  \right) \right]  
  \,,\label{kappaMF1}
  \\
  c_{SW}^{MF }& = & \frac{u_S}{u_0^4} 
  \Big\{ 1   +   g^2_{MF} \,
  \Big[
   C_F\,\left(-0.0211635 +  0.115961\,\omega +  0.685247\,\omega^2 \right)
  \nonumber\\
  & &+\, 
  N_c\,\left(0.013777 + 0.015905\,\omega - 0.321899\,\omega^2\right)\,\Big] 
  \Big\}\,.\label{cSWMF1}
\end{eqnarray} 
The mean field improved quantities are expressed in terms of the boosted coupling
$g^2_{MF}=g^2/u_0^4$. In~\cite{Horsley:2008ap} it is shown that $g^2_{MF}$ is
a good expansion parameter in the case of the Symanzik improved gauge action with
SLiNC fermions: in one-loop it does not differ very much from the coupling
$g^2_{\overline{MS}}$ in the $\overline{MS}$-scheme.
Comparing (\ref{kappaMF1}) and (\ref{cSWMF1}) with (\ref{kappacSym}) and (\ref{cswSym}) we  find
that the one-loop correction terms are indeed smaller than in the naive 
perturbative expressions. Therefore, the mean field approximation has improved
the perturbative behaviour as expected.

\section{Renormalization of currents}

We consider the renormalization constants for the following local bilinear
quark operators
\begin{equation}
S=\bar{\psi}{\rm \bf 1}\psi,\quad P=\bar{\psi}\gamma_5\psi,\quad V=\bar{\psi}\gamma_\mu\psi,\quad
A=\bar{\psi}\gamma_\mu\gamma_5\psi\,.
\end{equation}
The corresponding renormalization factor for an operator $O$ has the general form
\begin{equation}
Z_{O} = 1 - \ggCF\,\left(\gamma_{O}\,\log(a^2\mu^2) + \mathcal{B}_O(\omega)   \right) + \mathcal{O}(g^4)\,.
\label{ZO}
\end{equation}
Applying the mean field improvement as discussed in the preceding section the Z-factor
is obtained as
\begin{eqnarray}
Z_{O}^{MF} &=&  u_S\,\Big(1 - \frac{g^2_{MF}\,C_F}{16\pi^2}\,\left(\gamma_{O}\,\log(a^2\mu^2) + \mathcal{B}_{O}(\omega) -
	k_S(\omega)  \right)+ \mathcal{O}(g^4_{MF})\Big)\nonumber\\
&\equiv& u_S\,\Big(1 - \frac{g^2_{MF}\,C_F}{16\pi^2}\,\left(\gamma_{O}\,\log(a^2\mu^2) + \mathcal{B}_{O}^{MF}(\omega) \right)
 + \mathcal{O}(g^4_{MF})\Big) \,,
\label{ZOMF}
\end{eqnarray}
where $k_S(\omega)$ is given in (\ref{kS}).
We expect that stout smearing leads to a more continuum-like behavior. For the choice $a \approx 1/\mu$ this
means that the correction term $\mathcal{B}_{O}$ should become small in order to achieve
$Z_{O} \approx 1$. In Table \ref{tab1} we show the results for the local operators. It is
obvious that due to smearing with the selected value $\omega=0.1$~\cite{Horsley:2008} the one-loop
correction is diminished essentially.
\begin{equation}
\label{tab1}
\begin{tabular}{c|c|c|c|c|c|c}
O  & $\gamma_O$ & $\mathcal{B}_O(\omega)$ & $\mathcal{B}_O(0)$& $\mathcal{B}_O(0.1)$& $\mathcal{B}_O^{MF}(0)$&$\mathcal{B}_O^{MF}(0.1)$\\
  \hline
  &&&&&\\[-2ex]
  $S$    &  $-3$   & $15.075- 168.341 \omega + 242.254 \omega^2$  &$15.075$ & $0.663$  & $\phantom{1}7.845$& $-0.280$\\[0.7ex]
  $P$    &  $-3$    & $19.150- 267.462 \omega + 1065.55 \omega^2$  &$19.150$ & $3.059$ & $11.920$& $\phantom{-}2.117$\\[0.7ex]
  $V$    &  $\phantom{-}0$    & $11.911- 170.763 \omega + 754.029 \omega^2$  &$11.911$ & $2.375$& $\phantom{1}4.681$ & $\phantom{-}1.432$\\[0.7ex]
  $A$    &  $\phantom{-}0$   &  $10.717- 127.200 \omega + 342.380 \omega^2$  &$10.716$ & $1.420$& $\phantom{1}3.487$ & $\phantom{-}0.478$\\[0.7ex]
\end{tabular}
\end{equation}
In order to show the effect on the renormalization factors themselves we need the values for
$u_S$ and $u_0$. For  $\beta=5.5$ we have
$u_S=0.9404$ and  $u_0=0.8495$~\cite{Zanotti}. The following table shows the corresponding results
for $a=1/\mu$ and this selected $\beta$-value
\begin{equation}
\label{tab2}
\begin{tabular}{c|c|c|c}
O  & $Z_O(\omega=0)$ & $Z_O(\omega=0.1)$ & $Z_O^{MF}(\omega=0.1)$\\
  \hline
  &&&\\[-2ex]
  $S$       & $0.768$  &$0.990$ & $0.948$ \\[0.7ex]
  $P$       & $0.706$  &$0.953$ & $0.882$ \\[0.7ex]
  $V$       & $0.817$  &$0.964$ & $0.901$ \\[0.7ex]
  $A$       & $0.836$  &$0.978$ & $0.927$ \\[0.7ex]
\end{tabular}
\end{equation}
In (\ref{tab2}) we see that smearing shifts the renormalization factors towards unity 
showing a better continuum-like behaviour as promised.

\end{document}